\documentstyle[amssymb,pra,aps,psfig]{revtex}

\begin{document}
\draft

\twocolumn[\hsize\textwidth\columnwidth\hsize\csname @twocolumnfalse\endcsname
\title{Instability Heating of Sympathetically-Cooled Ions in a Linear Paul Trap}
\author{T. J. Harmon,  N. Moazzan-Ahmadi,  R. I. Thompson}
\address{Department of Physics and Astronomy, University of Calgary, 2500 University Drive NW,
Calgary, AB, Canada, T2N 1N4}
\date{\today}
\maketitle

\begin{abstract}

Sympathetic laser cooling of ions stored within a linear-geometry, radio 
frequency, electric-quadrupole trap has been investigated using computational and 
theoretical techniques.  The simulation, which allows 5 sample ions to interact 
with 35 laser-cooled atomic ions, revealed an instability heating mechanism, 
which can prevent ions below a certain critical mass from being sympathetically 
cooled.  This critical mass can however be varied by changing the trapping field 
parameters thus allowing ions with a very large range of masses to be 
sympathetically cooled using a single ion species.  A theoretical explanation of 
this instability heating mechanism is presented which predicts that the cooling-heating
boundary in trapping parameter space is a line of constant $q_u$ (ion trap 
stability coefficient), a result supported by the computational results.  The threshold value 
of $q_u$ depends on the masses of the interacting ions. A functional 
form of this dependence is given.

\end{abstract}
]

\narrowtext

\section{introduction}

\label{sec:introduction}

Laser cooling~\cite{1} has allowed atomic physicists to study gas-phase
atomic systems at ultra-low temperatures, largely removing thermal effects
and allowing many atomic systems to be studied at their most fundamental
quantum level (e.g. atom interferometry~\cite{2}). However, although a range
of neutral and ionic atomic systems has been cooled to sub-Kelvin
temperatures with direct laser cooling~\cite{1}, there is a limitation of
this approach. Conventional laser cooling techniques, both Doppler~\cite{3}
and sub-Doppler~\cite{4}, require isolated two- or three-level systems. If a
system has additional low-energy meta-stable states or multiple nearly
degenerate ground states (e.g.\ rotational structure in molecular ions) then
the laser-cooling cycle terminates when the system falls into a
non-laser-cooled level. The solution to this problem generally requires one
additional re-pumping laser system for each meta-stable state (e.g.\ Ref.~%
\onlinecite{5}). Cooling of molecules is especially challenging due to their
large number of accessible rotational states. Understandably, there only
exist a small number of very complicated theoretical proposals (e.g.\ Ref.~%
\onlinecite{6}) to directly cool molecular ions. Therefore, many atomic
systems such as those lacking the necessary energy level structure and
virtually all molecular systems cannot be directly laser cooled. However, a
much wider range of atomic and molecular systems can be taken to very low
temperatures through a process known as sympathetic cooling.

In this technique, a collection of two interacting species, the sample
species (which can not be directly laser cooled) and the laser-cooled
species, are confined to the same region of space and isolated from their
surroundings through the use of neutral particle or ion traps. As the
laser-cooled species interacts with the light field its temperature falls.
When these cold atoms interact or collide with the sample species, they
extract energy from the sample species thus sympathetically reducing the
sample species temperature. Recently, a variation of this approach has been
applied to neutrals whereby evaporatively-cooled atoms sympathetically cool
a sample species to produce Quantum Degenerate Gases from species whose
self-interaction (the interaction between one atom with another identical
atom) is too weak to permit efficient evaporative cooling on its own~\cite{7}%
. This type of sympathetic cooling in neutrals depends fundamentally on the
short range interactions (collisions) between the directly-cooled and sample
species, and thus the effectiveness of sympathetic cooling depends
fundamentally on the particular pair of species selected and their
associated inter-atomic or atom-molecule interaction potentials. This makes
it rather difficult to produce simple and broadly general rules regarding
the effectiveness of sympathetic laser cooling that would apply to all
neutral atomic and molecular species. However, when considering sympathetic
cooling between ions, the dominant interaction force is Coulomb
interactions, which are both long range and quite strong. This means that
ions do not tend to approach each other close enough that the short-range
interaction potentials, which differ from atom to atom and molecule to
molecule, become important. Therefore, the effectiveness of sympathetic
cooling between charged particles tends to be a function of only the mass
and charge of each of the two ions, allowing for rather generalised rules
regarding sympathetic cooling of charged particles to be deduced.

This work marks the beginning of a theoretical and experimental program to
study the processes associated with the sympathetic cooling of ions. The
longer term goals are to develop the tools and expertise to produce
low-temperature, gas-phase samples in as wide a range of atomic and
molecular species as is physically possible for use in such areas as
molecular spectroscopy, spectroscopy of stable and unstable atomic isotopes,
and the physical implementation of Quantum Information Science protocols.
The following pages outline a computational/theoretical study of the
apparently simple question: What is the range of ion masses that a
particular laser cooled ionic species can sympathetically cool in a
quadrupole ion trap?

Sympathetic cooling of molecular ions by laser-cooled atomic ions within a
linear Paul trap has been experimentally demonstrated~\cite{8} using ions of
similar mass; reaching translational temperatures below 100 mK. Further
experimental and molecular dynamics (MD) simulations have shown axial
translational temperatures as low as 10 mK using a single ion species to
sympathetically cool over a wider mass range~\cite{9}, but no particular or
general limits on this range were determined. More recently, work related to
a novel ion trap in-situ mass spectrometry technique under development by
the Baba and Waki group~\cite{10} has lead to an MD simulation study, which
illustrated that for particular trap conditions there is a lower bound on
the sample ion mass ($m_{s}$) that can be sympathetically cooled. This limit
is 0.54 times the mass of the laser cooled ion ($m_{c}$)~\cite{11}. However,
no upper bound on the masses that could be sympathetically cooled was
examined. This work investigates both the cooling of higher mass ions,
indicating a lack of a concrete upper bound, and examines the question of
sympathetic cooling of light sample ions in a more general setting. The
latter point lead to techniques to modify the cooling threshold value, and
provided a general theoretical framework for the heating processes in terms
of the concept of instability heating.

\section{Sympathetic Cooling of Trapped Ions: The Basics}

\label{sec:basics}

As mentioned in the previous section, sympathetic cooling of ions operates
through Coulomb interactions with laser cooled ions. Theoretically, such
interactions between ions that are completely isolated from their
surroundings, other than via the laser-cooling beam, will eventually bring
all of the ions into thermal equilibrium at the laser-cooled ion
temperature. However, trapped ions are not completely isolated because the
electric fields (and magnetic fields for Penning Traps) used to suspend the
ions in space can themselves produce a number of heating mechanisms (e.g.\
rf-heating~\cite{12}) and it is the balance of laser cooling and trap
heating that leads to an equilibrium temperature. Therefore, depending on
the relative efficiency of energy extraction through collisions with laser
cooled ions and trap related heating effects a given ion can either increase
or decrease its energy. It is this balance that leads to the heating-cooling
threshold observed in Ref. \onlinecite{11}. In this work, we will
concentrate on ions stored in a linear-geometry, rf quadrupole-electric
field trap (commonly referred to as a linear Paul trap). In this type of
trap, the ions are contained in the z-direction by a DC trapping field,
while a rf-oscillating saddle-shaped potential traps the ions in the
xy-plane~\cite{13}.

The stability of a single charged particle in an oscillating electric
quadrupole field is discussed in many references~\cite{13,14,15}.
Highlighting the relevant points, we start with a single ion in a linear
Paul Trap, where the confining potential along the z-axis of the quadrupole
is approximated as harmonic. The ion is subject to the following potential
in x and y coordinates

\begin{equation}
\Phi(x,y)={\frac{U_{rf}\left(y^2-x^2\right)\cos(\Omega t)}{2r_0^2}},
\label{phi}
\end{equation}

\noindent where $U_{rf}$ and $\Omega $ are the amplitude and angular
frequency of the applied rf-electric potential, respectively, and $r_{0}$ is
the distance from trap centre to one of the electrodes. Given that ${\bf F}%
=-e\nabla \Phi $, the equations of motion for an ion of mass $m$ and charge $%
e$ in the x-y plane are given by

\begin{eqnarray}
\ddot{x}+(e/mr_{0}^{2})U_{rf}\cos (\Omega t)x &=&0,  \nonumber \\
\ddot{y}-(e/mr_{0}^{2})U_{rf}\cos (\Omega t)y &=&0.  \label{motion}
\end{eqnarray}

\noindent Recasting these equations in unitless parameter form $(\xi=\Omega
t/2,q_u=q_x=-q_y=2eU_{rf}/m\Omega^2r_0^2)$ we get

\begin{equation}
{\frac{\partial^2 u}{\partial\xi^2}}-[2q_u\cos(2\xi)]u=0,  \label{mathieu}
\end{equation}

\noindent where $u$ represents either x or y. Equation~(\ref{mathieu}) is the
Mathieu equation in its canonical form \cite{19}, its general solution being

\begin{equation}
u=\alpha^{\prime}e^{\mu\xi}\sum_{n=-\infty}^{\infty}C_{2n}e^{2in\xi}
+\alpha^{\prime\prime}e^{-\mu\xi}\sum_{n=-\infty}^{\infty}C_{2n}e^{-2in\xi}
\label{u}
\end{equation}

\noindent where $\alpha ^{\prime }$ and $\alpha ^{\prime \prime }$ are
integration constants determined by the initial conditions, (i.e.\ $u(\xi
_{0})$, $\dot{u}(\xi _{0})$, and $\xi _{0}$), while $C_{2n}$ and $\mu $
depend solely on the value of $q_{u}$ and not the initial conditions~\cite
{14}. If $\mu $ is purely imaginary then the ion will have an oscillatory
motion (trapped particle). However, if $\mu $ is real or complex the motion
will be unbound (unstable particle). The determination of $\mu $ from $q_{u}$
in general is a difficult problem and is best handled numerically. It is
well known that if $|q_{u}|<0.908$, then $\mu $ is purely imaginary and if $%
|q_{u}|>0.908$ $\mu $ becomes complex, other than for some very localised
values of $q_{u}$~\cite{13}. For this reason $q_{u}$ is called the stability
parameter.

For stable trapping parameters, the motion of an ion in x and y, as
described by Eq.~(\ref{u}), can be broken down into a superposition of the
ion's micromotion and its secular motion. The micromotion is an oscillatory
motion at the angular frequency of the trapping field that increases in
amplitude the further the ion is from the trap centre. It is the direct
result of the oscillation of the trapping field driving the ion back and
forth as the field reverses. The lower frequency secular motion is a simple
harmonic oscillation in the time-averaged pseudo-potential generated by the
trapping field. The secular frequency of this motion, for small values of $%
q_{u}$, is given by~ \cite{14}

\begin{equation}
\omega_{{\rm sec}}={\frac{q_u}{\sqrt{8}}}\Omega,  \label{sec}
\end{equation}

\noindent while the amplitude can be expressed in terms of the temperature
based on the assumption of $\frac{1}{2}k_BT$ of energy in the relevant mode:

\begin{equation}
u_{0}={\frac{4}{q_{u}\Omega }}\sqrt{\frac{k_{B}T}{m}}.  \label{u0}
\end{equation}

Sympathetic cooling requires that multiple interacting ions are
simultaneously stored in the trap. We cannot deal with such a many-body
system analytically, but rather it is possible to calculate the motion of
each ion by integrating the equations of motion for each ion in the trap.
Additional terms have to be added to Eq.~(\ref{motion}) to account for the
Coulomb repulsion between the ions, as well as the laser cooling~ \cite{16}.
The equation of motion for z is similar to the x and y equations, although
it is simpler because the electric field in this direction is DC.

\begin{figure}[tbp]
\begin{center}
\psfig{figure=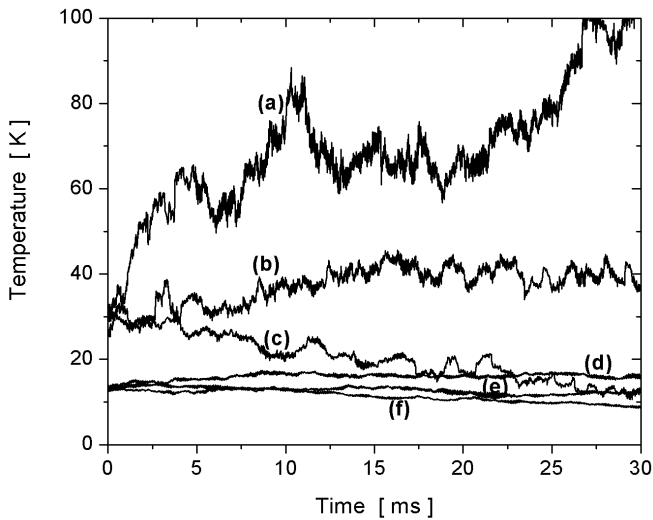,width=\columnwidth,angle=0}
\end{center}
\caption{Temporal plots of the average temperature (see text) of 8 amu
sample ions and 24 amu laser-cooled ions stored in a linear Paul trap
calculated under three different trapping conditions. Line (a) is the sample
ion temperatures for $U_{rf}=50$~V, $\Omega =2\protect\pi $(2.5 MHz), and $%
q_{u}=0.71$; line (b) is the sample ion temperatures for $U_{rf}=100$~V, $%
\Omega =2\protect\pi $(4.5 MHz), and $q_{u}=0.53$; line (c) is the sample
ion temperatures for $U_{rf}=30$~V, $\Omega =2\protect\pi $(2.75 MHz), and $%
q_{u}=0.35$. Lines (d), (e), and (f) are the laser-cooled ion temperatures
corresponding to the conditions for (a), (b), and (c), respectively.}
\label{1}
\end{figure}

Our simulation numerically integrated the motion of 35 laser cooled Mg$^{+}$
ions interacting with 5 sample ions using the Euler-Picard
predictor-corrector method with 0.1 to 1 ns time-steps. The ions are
confined radially by the potential given by Eq.~(\ref{phi}) and axially by a
harmonic well with a force constant of 1 mV/mm. The initial conditions for
the simulations started the ions in random positions with the sample ions at
a temperature of 40 K and the laser cooled ions at a temperature of 15 K.
The laser intensity was selected so that in the absence of sample ions,
rf-heating balanced the laser cooling and the laser-cooled ion temperature
would remain the same. In addition, the laser intensity was maintained at a
low enough level to prevent the formation of Wigner or ion crystals. These
conditions were chosen to be equivalent to those used in Ref.~\onlinecite{11}%
. The simulation output provided the average sample ion and laser-cooled ion
temperatures as a function of time. The temperature was calculated by
averaging the energy of each ion over 4 cycles of secular motion and then
calculating the average ion energy for the 5 sample ions and the 35 laser
cooled ions, independently. The ion temperatures were determined from these
average energies using~\cite{11}

\begin{equation}
\langle E\rangle={\frac{5}{2}}k_BT.
\end{equation}

Integration was generally carried out for 30 ms, which was usually
sufficient to determine whether the temperature of the sample ions was
increasing or decreasing with time. First, the laser-cooled ions were
interacted with a range of more massive sample ions (up to 8 times the mass
of the laser-cooled ions). This confirmed that sample ions heavier than the
laser-cooled ions generally cooled. However, as the mass difference
increased, the rate of cooling decreased.

\begin{figure}[tbp]
\begin{center}
\psfig{figure=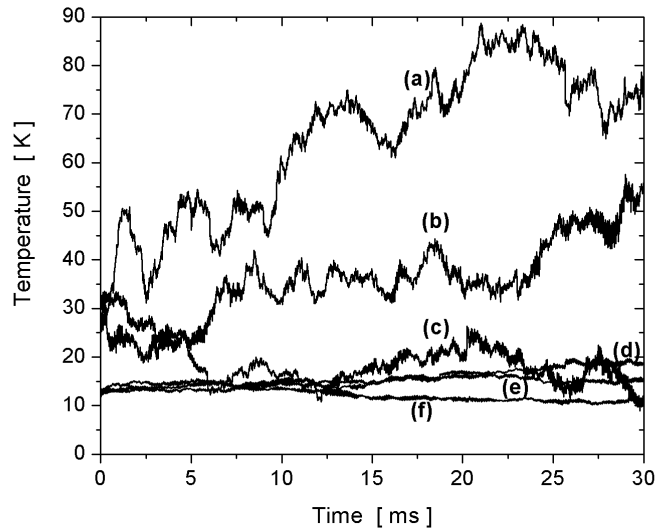,width=\columnwidth,angle=0}
\end{center}
\caption{Temporal plots of the average temperature (see text) of 12 amu
sample ions and 24 amu laser-cooled ions stored in a linear Paul trap
calculated under three different trapping conditions. Line (a) is the sample
ion temperatures for $U_{rf}=40$~V, $\Omega =2\protect\pi $(2 MHz), and $%
q_{u}=0.59$; line (b) is the sample ion temperatures for $U_{rf}=70$~V, $%
\Omega =2\protect\pi $(2.6 MHz), and $q_{u}=0.61$; line (c) is the sample
ion temperatures for $U_{rf}=40$~V, $\Omega =2\protect\pi $(2.5 MHz), and $%
q_{u}=0.35$. Lines (d), (e), and (f) are the laser-cooled ion temperatures
corresponding to the conditions for (a), (b), and (c), respectively.}
\label{2}
\end{figure}

Next the case in which the sample ions are less massive than the
laser-cooled ions was examined. For a given $U_{rf}$ and $\Omega $, heating
or cooling of the sample ion was observed to depend on its mass. We
confirmed that for the same trap conditions as in Ref.~\onlinecite{11},
there is the same lower bound on the sample ion mass that can be
sympathetically cooled, i.e.\ $m_{s}=0.54m_{c}$. At this point, tests were
made to determine if sample ions with $m_{s}<0.54m_{c}$ could be cooled by
changing the trap parameters. It was found that there always exist values of 
$q_{u}$ for which this is possible. A large number of simulations for a
range of sample ions with mass below $0.54m_{c}$ were carried out. In all
cases the results were more or less the same as those illustrated in Figs.~%
\ref{1} and \ref{2}. Figure~\ref{1} is for ($m_{s}=8$ amu, $m_{c}=24$ amu)
and Fig.~\ref{2} is for ($m_{s}=12$ amu, $m_{c}=24$ amu). Both of these
figures show the typical result that if the value of $q_{u}$ is reduced
sufficiently, the sample ions can be cooled. To further test this
hypothesis, maps of the $U_{rf}$-$\Omega $ space were made to determine the
trapping field parameters for which the sample ions heated, cooled, or
remained basically unchanged in temperature. Figure~\ref{3} shows a pair of
typical maps. The $q_{u}=0.908$ curve indicates the single sample ion
trapping stability threshold. As can be seen, sympathetic cooling does not
occur close to this threshold and thus the ions are heated. As we move away
from this threshold we enter a region where there appears to be a balance
thus the ions neither heat nor cool. Finally, for low enough values of $q_{u}
$ sympathetic cooling begins to occur. This begs the question of why does
cooling occur away from the single ion stability threshold and can we
predict the threshold for sympathetic cooling? The answer to these questions
are given in the next two sections.

\section{Instability Heating Theory}

\label{sec:extended}

\begin{figure}[tbp]
\begin{center}
\psfig{figure=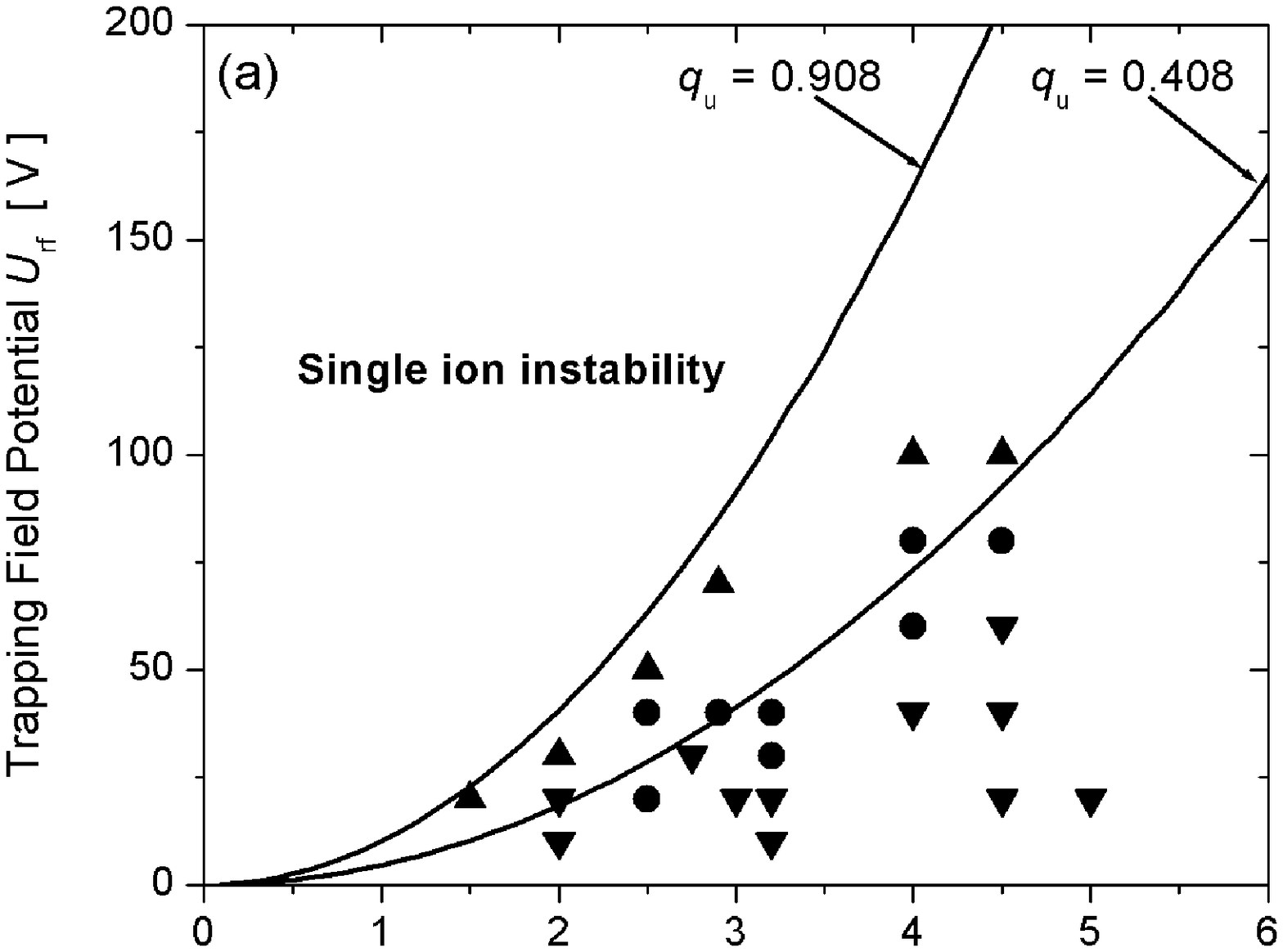,width=\columnwidth,angle=0}
\psfig{figure=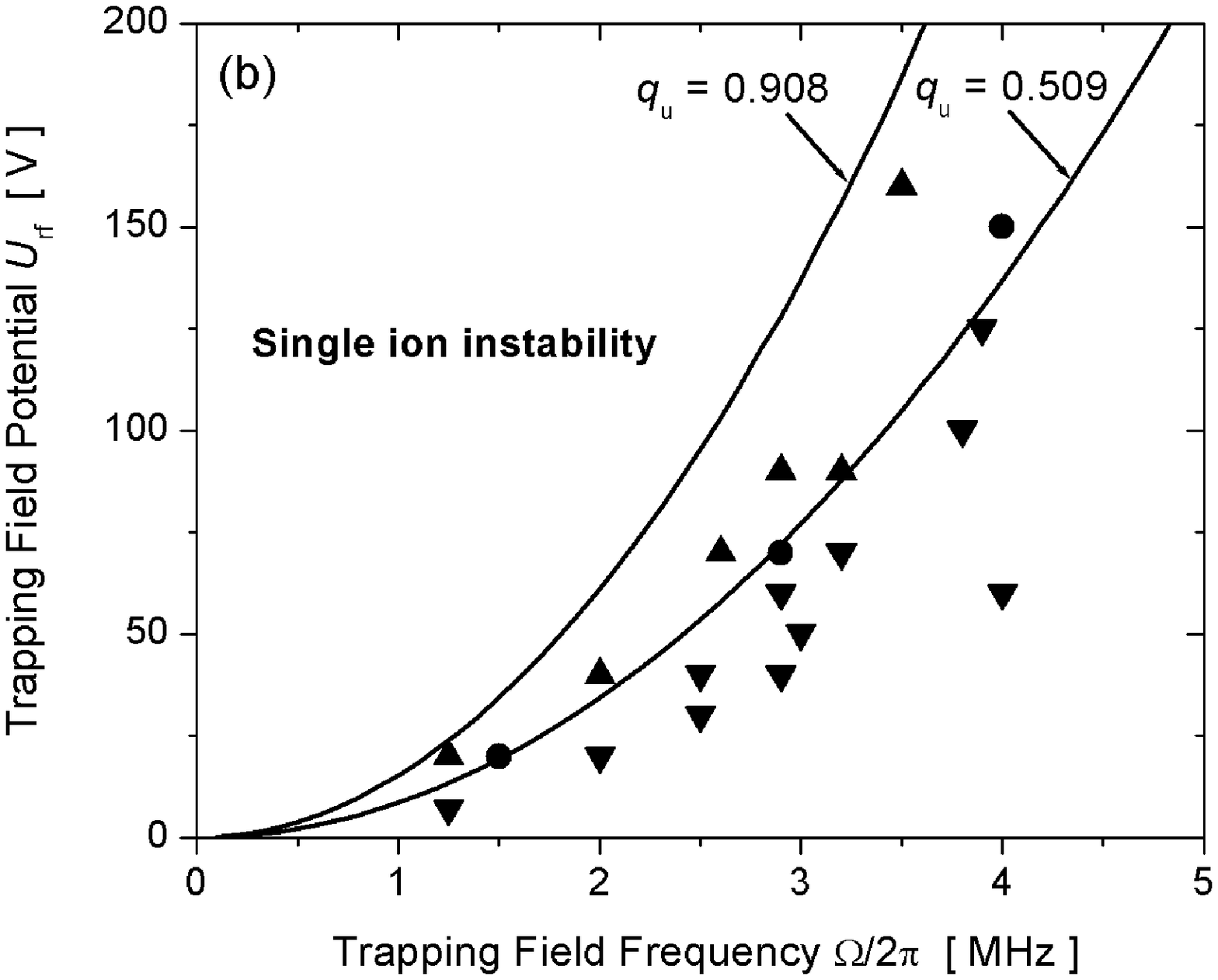,width=\columnwidth,angle=0}
\end{center}
\caption{Maps of the regions of heating and cooling in $U_{rf}-\Omega$ space
for 5 sample ions and 35 laser-cooled ions in a linear-geometry Paul trap. $%
\bigtriangleup$ indicates simulations that produced ion heating,
$\bigtriangledown$ indicates ion cooling, and $\bullet$
indicates no definitive change in temperature. Plot (a) is for 8 amu sample
ions and 24 amu laser cooled ions while plot (b) is for 12 amu sample ions
and 24 amu laser cooled ions. The solid line indicates the $q_u = 0.908$
stability threshold for single ion trapping, while the second constant $q_u$
line in each figure is the theoretical threshold between regions of
sympathetic cooling and heating of the sample ions (see text).}
\label{3}
\end{figure}

Although we have not yet discussed the precise heating mechanism
responsible, the results summarised in Fig.~\ref{3} clearly suggest the
existence of a second mass threshold for the behaviour of trapped ions. The
first mass threshold, dictated by the $|q_{u}|<0.908$ trap stability
criterion, determines if the ion stays in the trap. The second higher-mass
threshold, which is also a function of the trapping field parameters,
determines whether or not the sample ion can be sympathetically cooled.

From the results of the previous section, it is reasonable to conclude that
the two mass thresholds have similar functional dependences on the trapping
field parameters and ion mass, suggesting that the physical causes of these
two thresholds might be related. These considerations prompted us to
investigate if the heating process is related to transient instabilities in
the ion motion due to interactions with the other ions in the trap, an
effect not considered in the single ion treatment. We will refer to this
process as instability heating.

The basic physical mechanism for instability heating is as follows. 
Equation~(\ref{u}) gives the general form for motion of an ion in an oscillating
quadrupole field. For a single trapped ion, $q_{u}$ is a constant and for $%
|q_{u}|<0.908$, $\mu $ is purely imaginary thus the orbit is oscillatory.
When $|q_{u}|>0.908$, $\mu $ will contain a real part which will cause the
trajectory of the ion to grow exponentially. However, when the ion in
question is in the presence of other ions, $q_{u}$ can no longer be
considered a constant but the additional randomly varying Coulomb forces
require it to be replaced with a time-varying effective stability parameter $%
q_{{\rm eff},u}$. Viewed as such, $|q_{{\rm eff},u}|$ can now exceed the
threshold value of 0.908 for short periods of times without the ion escaping
the trap. These time intervals in which the ion trajectory grows
exponentially are followed by intervals for which $|q_{{\rm eff},u}|$ drops
below 0.908 thus re-stabilizing the ion at a larger maximum radius and a
higher energy orbit.

To calculate $q_{{\rm eff},u}$, we will consider our physical system at a
particular point in time and replace the sample ion of interest, which is
exposed to the trap forces ($F_{T,u}$) and Coulomb force due to other ions ($%
F_{i,u}$) present in the trap, with an equivalent ion. Furthermore, we will
assume that this equivalent ion is positioned at the same point in the trap
as the ion of interest, having the same velocity and acceleration, and is
subject to the same trap force but is not exposed to the Coulomb force due
to the other ions, thus requiring it to have a time-varying effective mass.
The effective mass of the equivalent ion in terms of the mass of the sample
ion is given by

\begin{equation}
m_{{\rm eff},u}={\frac{m}{\left(1+{\frac{F_{i,u}}{F_{T,u}}}\right)}}.
\end{equation}

\noindent Since the effective mass is the mass for an equivalent single ion
in a trap, Eq.~(\ref{mathieu}) can be used to obtain $q_{{\rm eff},u}$ given by

\begin{equation}
q_{{\rm eff},u}=q_u\left(1+\varepsilon_u\right),
\end{equation}

\noindent where

\begin{equation}
\varepsilon_u={\frac{F_{i,u}}{F_{T,u}}}.  \label{epsilon}
\end{equation}

\noindent Note that $q_{{\rm eff},u}$ is both time varying and can have
different values for the x and y directions. We can now write the
instantaneous stability criterion as

\begin{equation}
|q_{{\rm eff},u}|=\left|q_u\left(1+\varepsilon_u\right)\right|\leq 0.908.
\label{qeffu}
\end{equation}

\begin{figure}[tbp]
\begin{center}
\psfig{figure=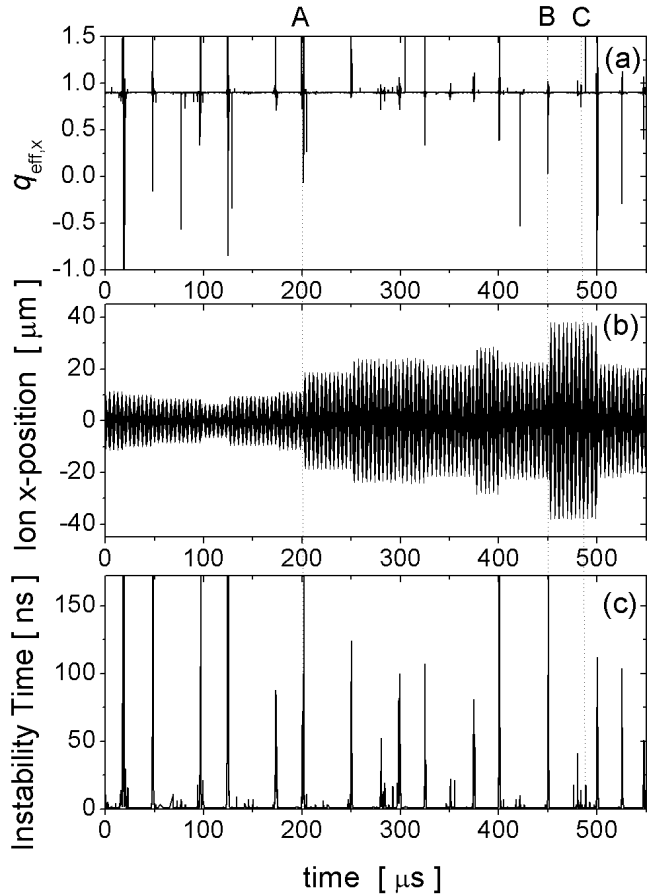,width=\columnwidth,angle=0}
\end{center}
\caption{Temporal plots of the computed evolution of (a) $q_{{\rm eff},x}$
(see text), (b) the x-position and (c) the time period that $q_{{\rm eff},x}$
remained above 0.908 for the sample ion in a simulation involving a 6.6 amu
sample ion and a 30 amu laser-cooled ion. The trapping field parameters were 
$U_{rf} = 70$~V, $\Omega = 2\protect\pi$(2.9 MHz), and $q_u = 0.9$.}
\label{4}
\end{figure}

When $\left| q_{{\rm eff},u}\right| >0.908$ (i.e.\ when the inter-ionic
forces sufficiently exceed the trapping forces) we would expect the
possibility of exponential growth of ion trajectory, leading to increased
ion temperatures. Figures~\ref{4}a and \ref{4}b illustrate that this is
indeed the case. These simulations show a short snapshot of the temporal
evolution of the x-coordinate of a sample ion in parallel with the computed
value of $q_{{\rm eff},x}$. They were obtained with a single sample ion, a
single laser-cooled ion, and with $q_{u}$ ($=0.9$) set very near the single
ion instability threshold in order to maximize the frequency of instability
points. The evolution of the particle motion shows a sequence of steps in
the amplitude of the ion's motion, corresponding to steps in the temperature
of the ion. For each of these steps, we observe a corresponding spike in $q_{%
{\rm eff},x}$, where $|q_{{\rm eff},x}|$ exceeds $0.908$. In some cases, the
amplitude steps down, which is expected since Eq.~(\ref{u}) contains both
exponentially growing and decaying terms. The energy drops only when the
decaying term is much larger than the growth term, while it increases under
all other conditions. On the other hand, there are definitely many points
where $q_{{\rm eff},x}$ exceeds $0.908$, but no significant change in the
motional amplitude occurs. The explanation for this curious behaviour is
given below.

What is perhaps more important than $q_{{\rm eff},u}$ merely exceeding 0.908
is the time interval over which $q_{{\rm eff},u}$ has an unstable value. As
one can see from Eq.~(\ref{epsilon}), even in the presence of a distant laser
cooled ion $\varepsilon _{u}$ undergoes a singularity each time the trapping
force on the sample ion becomes zero. This occurs at least twice per cycle
of the rf-trapping potential. However, when the time period over which $q_{%
{\rm eff},u}$ takes an unstable value is short compared to the time interval
for the rest of the cycle, the ion trajectory does not significantly grow,
hence no measureable heating occurs. Since the ion motion evolves at the
secular frequency, a singularity lasting for such a short time interval
should not significantly effect the energy of the sample ion or its
temperature.

\begin{figure}[tbp]
\begin{center}
\psfig{figure=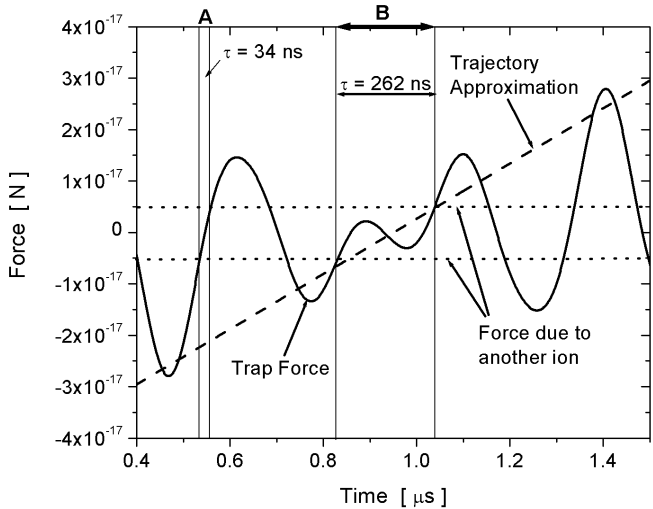,width=\columnwidth,angle=0}
\end{center}
\caption{A temporal plot of the computed trap force on a moving ion. Point A
illustrates the short instability period when the inter-ionic force exceeds
the trapping force, while period B illustrates the longer instability period
occurring when the secular motion of the ion passes through $x=0$.}
\label{5}
\end{figure}

More significantly, $\varepsilon _{x}$ (or $\varepsilon _{y}$) also
undergoes a singularity as the ion crosses the y (or x) axis. This occurs
for each half cycle of secular motion. The important difference between this
and the trapping field singularities is that the instability periods can
last over a much longer time period as illustrated in Fig.~\ref{5}. Here,
the solid curve represents the trap force on a moving ion as a function of
time and the horizontal dotted lines represent the magnitude of the slowly
varying Coulomb force due to another ion. For times less than 0.8 $\mu $s,
the amplitude of the trap force is large (i.e.\ the sample ion is not near
the y axis) and thus the inter-ionic force dominates only for very short
time-intervals as illustrated by A, which is too short to significantly
effect the trajectory. However, when the ion is close to the y axis (i.e.\
times near 1 $\mu $s), the trapping force is weaker and the inter-ionic
force dominates over a longer time period, possibly over several cycles of
the trapping field if the ion spacing is small. This is illustrated by event
B. This results in a measureable heating of the sample ion. The importance
of the period over which the ion remains unstable is shown in part (c) of
Fig.~\ref{4}. We see that the quantized steps in the amplitude of the ion
motion occur when $q_{{\rm eff},x}$ exceeds $0.908$ and does so for a period
exceeding 30 to 40 ns. To illustrate this point, three events are
highlighted. At point A, $q_{{\rm eff},x}$ spikes to well above 1 for well
over 100 ns, resulting in a clear step in the motional amplitude. At point
B, there is also an amplitude step, but here $q_{{\rm eff},x}$ barely
exceeds 1. However, $q_{{\rm eff},x}$ stays above the stability threshold
for a relatively long time ($>100$ ns). On the other hand, at point C there
are several spikes all exceeding 1. However, none of the spikes last longer
than 20 ns, and the ion motion shows no significant change.

\section{Calculation of Heating-Cooling Threshold}

To generate a predicted shape for the cooling-heating threshold, we need to
calculate a balance between the heating and cooling rates. \ To calculate
the heating rate, we first need the variation of the mean instability time
with the trapping field parameters. From Eq.~(\ref{epsilon}) and \ref{qeffu},
the time interval for which $|q_{{\rm eff},u}|$ is larger than 0.908 is
given by the time interval when

\begin{equation}
|F_{T,u}|\leq {\frac{|\left\langle F_{i,u}\right\rangle |q_{u}}{0.908\pm
q_{u}}}.  \label{Ftu1}
\end{equation}

\noindent is satisfied. The upper sign is used when $F_{T,u}$ and $F_{i,u}$
have opposite sign and the lower sign when they are pointed in the same
direction. $\left\langle F_{i,u}\right\rangle $ is the average inter-ionic
force during a collision. An estimate of this time interval can be obtained
as a function of trapping field parameters. The trapping force can be
determined by taking the gradient of Eq.~(\ref{phi}). Using Eqs.~\ref{sec} and 
\ref{u0}, the force on an ion near the trap centre can be approximated by
(see Fig.~\ref{5})

\begin{equation}
F_{T,x}={\frac{ex}{r_{0}^{2}}}U_{rf}\sin \Omega t^{\prime }={\frac{%
eu_{0}\sin \omega _{{\rm sec}}t}{r_{0}^{2}}}U_{rf}\sin \Omega t^{\prime },
\end{equation}

\noindent where $t$ and $t^{\prime }$ are used to indicate that the secular
motion and micromotion may be out of phase. We neglect the small effect of
micromotion on the x-coordinate of the ion near the trap axis. Given that
the time scale of interest is the secular period, much longer than the
trapping field period, the trapping field oscillation can be averaged out.
Strictly speaking this is zero, but stability is dependent on $|q_{{\rm eff}%
,u}|\leq 0.908$. Thus it is really the absolute value of the forces, not
their signs which is of interest. Therefore, it is reasonable to replace $%
\sin \Omega t^{\prime }$ with its rms value. Assuming that we are only
interested in times close to the point where the ion crosses the trap axis,
the small angle approximation $\omega _{{\rm sec}}t\ll 1$ can be used. Thus

\begin{equation}
F_{T,x}={\frac{eu_{0}\omega _{{\rm sec}}t}{2r_{0}^{2}}}U_{rf}.
\end{equation}

\noindent Using Eqs.~\ref{sec} and \ref{u0} , we find

\begin{equation}
F_{T,x}={\frac{eU_{rf}}{2r_{0}^{2}}}\sqrt{\frac{2k_{B}T}{m_{s}}}t.
\label{Ftu2}
\end{equation}

\noindent Substituting Eq.~(\ref{Ftu2}) into Eq.~(\ref{Ftu1}), and solving for
the instability time $\tau$, we find:

\begin{equation}
\tau ={\frac{0.908|F_{i,u}|q_{u}}{0.908^{2}-q_{u}^{2}}}{\frac{r_{0}^{2}}{%
eU_{rf}}}\sqrt{\frac{m_{s}}{2k_{B}T}}.  \label{tau}
\end{equation}

\noindent One can see immediately that the instability increases in duration
as $q_{u}$ approaches 0.908, and eventually becomes infinity at the point
that $q_{u}$ reaches the single ion instability threshold. 

Given Eq.~(\ref{tau}), we can calculate the heating rate by also considering
the collision rate, the frequency at which these axis-crossing instabilities
happen, and the rate at which the energy grows during the instability. \ The
collision rate is simply a function of the number of trapped ions and
trapping volume, while axis crossings occur at twice the secular frequency.
\ To determine the rate of energy growth during the instability, we must
calculate how rapidly the radius of the ion trajectory grows. The parameter $%
\mu $ in Eq.~(\ref{u}) governs this growth. Since $q_{u}=0.908$ marks the
transition from complex to real values for $\mu $, this coefficient, at
least near the transition point, should take on the form

\begin{equation}
\mu=\eta\sqrt{|q_u|-0.908},  \label{mu}
\end{equation}

\noindent where $\eta $ is a constant of order unity. This cannot be shown
analytically for the Paul trap. However, for the closely related rotating
saddle trap that can be solved analytically, $\mu $ takes the form of 
Eq.~(\ref{mu})~\cite{17}. Converting back from unitless parameters, the rate of
exponential growth in the motion during the instability is given by

\begin{equation}
{\frac{1}{2}}\mu \Omega ={\frac{1}{2}}\eta \Omega \sqrt{|q_{{\rm eff}%
,u}|-0.908},  \label{heat}
\end{equation}

\noindent while the rate of energy increase is simply twice this amount due
to the harmonic nature of the potential. \ Here, $q_{u}$ has been replaced
with $q_{{\rm eff},u}$ since it is the quantity that reaches unstable
values. We will approximate Eq.~(\ref{heat}) using a mean value for $q_{{\rm %
eff},u}$ by assuming that $F_{i,u}$ is roughly constant over the instability
period and that $F_{T,u}$ can be replaced by the average of its maximum
magnitude during instability and its minimum value of zero (see Eq.~(\ref
{Ftu1})). When $q_{{\rm eff},u}$ for parallel and anti-parallel forces are
averaged, the result is independent of $q_{u}$ leaving an average value for
the exponential energy growth rate of

\begin{equation}
\left\langle \mu \Omega \right\rangle =\eta \Omega \sqrt{0.908}.
\end{equation}

\noindent Finally, taking the heating rate to be proportional to the the
instability time, energy growth rate, the collision rate ($\sigma_{\rm coll}$),
and the axis crossing frequency ($2\sigma _{sec}$) yields

\begin{equation}
R_{h}={\frac{K_{h}\sigma_{\rm coll}0.908|F_{i,u}|q_{u}}{0.908^{2}-q_{u}^{2}}}{%
\frac{\eta }{4\pi }}\sqrt{\frac{0.908}{m_{s}k_{B}T}}.  \label{Rh}
\end{equation}

To avoid overall heating of the ions, $R_{h}$ must be balanced by the
cooling effect, which occurs via elastic collisions between slow and fast
moving particles of different masses. Simple classical mechanics arguments
show that the efficiency of elastic collision energy transfer decreases as
the mass difference between the particles increases. The energy transfer
efficiency is given by $4\rho /(1+\rho )^{2}$, where $\rho =m_{s}/m_{c}$~
\cite{10}. Taking the cooling rate to be proportional to the product of this
efficiency and the collision rate gives

\begin{equation}
R_{c}=K_{c}\sigma_{\rm coll}{\frac{m_{s}/m_{c}}{\left( 1+{\frac{m_{s}}{m_{c}}}%
\right) ^{2}}}.  \label{Rc}
\end{equation}

\noindent When the two rates balance out, we should see a transition from
heating to cooling of the sample ions. Equating Eqs.~\ref{Rh} and \ref{Rc},
we obtain

\begin{equation}
{\frac{q_{u}}{0.908^{2}-q_{u}^{2}}}={\frac{4\pi K}{0.908|F_{i,u}|\eta }}%
\sqrt{\frac{k_{B}T}{0.908}}{\frac{\sqrt{m_{s}^{3}}/m_{c}}{(1+m_{s}/m_{c})^{2}%
}},  \label{equil}
\end{equation}

\noindent where the constant K, which is independent of the trapping
parameters, is defined to be $K_{c}/K_{h}$. \ The collision rates in Eqs. 20
and 21 have been taken to be the same. \ This is valid if the number of
laser-cooled ions significantly exceeds the number of sample ions. In
general, cooling occurs when sample ions collide with laser-cooled ions,
while instability heating is caused by interactions between any two ions.
However, if most of the ions in the trap are laser-cooled ions, then most of
the instability heating is a result of interactions between sample ions and
laser-cooled ions. This means that both the heating and cooling rates are
proportional to the same collision rate, and thus this rate cancels out.
Therefore, the rate of collisions determines the rate of heating or cooling
and not the sign of the temperature change.

Equation~(\ref{equil}) provides the theoretical justification that lines of
constant $q_{u}$ define the heating-cooling boundaries in U$_{rf}$-$\Omega $
space. Although we cannot calculate threshold values for $q_{u}$ from first
principles, because of the unknown scaling factors, it can be done
numerically. This was done to produce the $q_{u}=0.509$ curve in Fig.~\ref{3}%
b, showing that this line effectively separates the regions of heating and
cooling. Once the threshold value of $q_{u}$ is determined for one mass
combination, Eq.~(\ref{equil}) can be used to determine the value for other
mass combinations. This is illustrated by the $q_{u}=0.408$ line in Fig.~\ref
{3}a. This line was not fit to the data, but rather it was calculated from
the constant value for Fig.~\ref{3}b. It should be noted that this model has
been applied only to a limited mass range and that perhaps other mass
effects have to be included if very large mass differences are to be
considered.

Up to this point, we have concentrated mainly on the effect of the mass of
the sample ion. However, the laser-cooled ion mass is also significant. This
is because the efficiency of energy transfer between two colliding particles
decreases as the ratio of the masses deviates from unity, becoming
effectively zero when one mass is negligible as compared to the other.
However, the instability-heating rate is expected to be relatively unchanged
as it results from the Coulomb interactions between the ions and with the
trapping field. Thus for larger mass ratios, the heating-cooling threshold
should move further from the single-ion instability threshold in order to
reduce the amount of instability heating that must be overcome. This reduces
the overall collision rate which decreases the rate of heating or cooling of
the sample ion. The effect of laser-cooled ion mass was examined by running
simulations with 18 amu sample ions interacting with either 24 amu or 36 amu
laser cooled ions for a range of $q_{u}$. The results are summarized in
Table I which illustrates the expected slight shift in the cooling-heating
threshold for different laser-cooled ion masses. These results are in good
agreement with the heating-cooling transition values predicted by Eq.~(23)
based on the 12 amu/24 amu threshold. The predicted heating-cooling
threshold values are 0.585 for 18 amu/24 amu and 0.56 for 18 amu/36 amu.

\section{Discussion and Future Directions}

The results of the preceding sections provide us with a physical insight
into the competing processes that lead to sympathetic heating or cooling of
ion mixtures. Generally speaking, we can conclude that if sympathetic
heating is occurring for light sample ions, the trap parameters can be
adjusted to move away from the single ion instability threshold, thereby
allowing the sample ions to cool. However, there is a limit to this effect.
As one moves far away from the instability threshold, as would be needed to
cool very light ions, the trapping field becomes weaker, the trapping volume
grows, and the collision rate becomes very small. Although an ion still
cools, the rate of cooling may be so small as to be of little practical use.

Instability heating is to be distinguished from rf-heating \cite{13,18}. \
This transient effect is caused by random interactions between ions,
temporarily driving them into unstable motion. \ In contrast, rf-heating of
theions near the quadrupole axis is caused by continuous Coulomb
interactions with the ions which are constantly driven by rf field at the
periphery. Rf-heating cannot be suppressed solely by changing the trap
parameters, only by reducing the numbers of ions and cooling the sample to
the point where all the ions crystallize along the central axis of the trap
where micromotion is not present. However, instability heating, which is a
major source of heating in the low-density gas phase regime, can be quenched
by choosing favourable trap parameters so that the ions are less susceptible
to heating. This is achieved by minimising the duration of potentially
unstable motion and the frequency at which these potential instabilities
occur, so that in the event of a collision the ions is more likely to
decrease its energy.

Further theoretical work is needed to study the effect of changing the
relative numbers of sample and laser cooled ions. When the number of sample
ions increases, the interactions between them can no longer be ignored. This
will ultimately cause more heating, since such interactions cannot result in
a decrease in the mean energy. Although, Eq.~(16) suggests that instability
heating should decrease in the higher energy regime this has yet to be
tested. Such collisions at high temperatures ($\thicksim $300K) require much
more computational time and other techniques may be developed to model them.
At low energies ($\thicksim $1K) Eq.~(16), which is applicable for ions in the
gas phase, becomes less valid since the ions are in the crystalline phase.
Instability heating is expected to be insignificant here since the axial
separation is always large enough to prevent the transverse forces, which
cause instabilities, from becoming too large. The dominant heating mechanism
in this low energy crystalline phase would be rf-heating when the number of
ions is large.

In this work, we have ignored the internal degrees of freedom of the sample
ions, which is reasonable if the ions are all atomic, but less so in the
case of molecular sample ions. The distinction between atomic sample ions
and molecular sample ions was deliberately overlooked, since the focus of
this work is the translational degrees of freedom. In ignoring the molecular
ions internal structure, we have a first order approximation as to whether
or not the sample molecular ions will translationally heat or cool.
Translational cooling in this approximation is seen as a significant
component of a complete model of the sympathetic cooling. This is because a
major energy transfer process is expected to be intra-molecular transfer of
energy from rotational and vibrational modes to translational degrees of
freedom followed by inter-ionic translational energy transfer to the
laser-cooled atoms, as modeled in this work.

\section{Conclusions}

The mechanism through which sample ions are heated in a linear Paul trap is
a direct result of collisions, the same collisions that would otherwise
bring about thermal equilibrium at a lower temperature. A collision can
cause an ion to become briefly unstable and thus rather than energy being
transferred to the laser cooled ion for dissipation in the radiation field,
kinetic energy is pumped into the ion from the electric fields if its $%
q_{{\rm eff},u}$ is greater than 0.908. We have shown that instability heating 
is minimised when $q_{u}$ is small. Since $q_{u}$ is a function of mass, trap
frequency, and trap depth, although a lower mass limit exists below which
sample ions will heat, this threshold can be changed by modifying the trap
parameters. This mass limit corresponds to a second characteristic threshold
value for $q_{u}$, above which heating will occur \cite{11}. More generally,
there exists a three dimensional space spanned by mass, trap frequency, and
trap amplitude which is divided into regions of sympathetic heating-cooling.
Constraining any two of these variables will result in an upper or lower
limit in the third.

\section{Acknowledgments}

The Natural Sciences and Engineering Research Council of Canada supported
this research. One of the authors (TH) would like to acknowledge the
financial support of the Faculty of Graduate Studies in the University of
Calgary. D. Hobill and H. Graumann provided very useful assistance related
to the challenges of computational physics. The authors thank D. Feder for
his helpful criticism and assistance in preparation of this manuscript.

\begin{table}[tbp]
\begin{tabular}{|c|c|c|c|c|}
\hline
$q_{u}$ & $\Omega /2\pi $ [MHz] & $U_{rf}$ [V] & $m_{c}=24$ amu & $m_{c}=24$
amu \\ \hline
0.19 & 2 & 70 & C & C \\ \hline
0.25 & 2.5 & 40 & C & C \\ \hline
0.35 & 1.5 & 20 & C & C \\ \hline
0.38 & 2.5 & 60 & C & C \\ \hline
0.39 & 2 & 40 & C & C \\ \hline
0.43 & 2.54 & 70 & C & C \\ \hline
0.47 & 2.43 & 70 & C & N \\ \hline
0.51 & 2.5 & 80 & N & N \\ \hline
0.59 & 2 & 60 & N & N \\ \hline
0.61 & 3 & 140 & N & H \\ \hline
0.63 & 2.5 & 100 & N & H \\ \hline
0.66 & 2.05 & 70 & N & H \\ \hline
0.70 & 1.5 & 40 & H & H \\ \hline
0.76 & 2.5 & 120 & H & H \\ \hline
\end{tabular}
\caption{Computational observations of sympathetic heating or cooling for
different trapping field parameters for 5 sample ions ($m_{c}=18$ amu)
interacting with 35 laser cooled ions (either $m_{c}=24$ amu or $m_{c}=36$
amu). The sample ion behavior was categorized, based on a 60 ms evolution,
into heating (H), cooling (C), or no observed temperature change (N).}
\label{ptable}
\end{table}

\end{document}